\documentclass[twocolumn,prd,showpacs,nofootinbib,superscriptaddress,floatfix]{revtex4}
\usepackage{latexsym}
\usepackage{dcolumn}
\usepackage{color}
\RequirePackage{graphicx}
\usepackage{epsfig}
\newcommand{\ba}{\begin{eqnarray}}

\newcommand{\ea}{\end{eqnarray}}
\newcommand{\be}{\begin{equation}}
\newcommand{\ee}{\end{equation}}
\newcommand{\beq}{\begin{equation}}
\newcommand{\eeq}{\end{equation}}
\newcommand{\beqarray}{\begin{eqnarray}}
\newcommand{\eeqarray}{\end{eqnarray}}
\newcommand{\bdisplay}{\begin{displaymath}}
\newcommand{\edisplay}{\end{displaymath}}

\newcommand{\calF}{{\cal F}}

\newcommand{\eq}[1]{Eq.\,(\ref{#1})}

\newcommand{\greaterabout}{\,\raisebox{-.6ex}{\ $\stackrel{>}{\sim }$\ }\,}

\newcommand{\hatF}{\mbox{${\hat {\cal F}}(v,Q^2)$}}

\def\bea{\begin{eqnarray}} 
\def\eea{\end{eqnarray}} 
\newcommand{\la}{\,\raisebox{-.8ex}{\,$\stackrel{\textstyle <}{\sim}$}\,\,} 
\newcommand{\ga}{\,\raisebox{-.8ex}{\,$\stackrel{\textstyle >}{\sim}$}\,\,}

\newcommand{\asoverpi}{{\alpha_s\over 4\pi}}

\begin{document}

\title{Analytic treatment of leading-order parton evolution equations:  theory and tests }

\author{Martin~M.~Block}
\affiliation{Department of Physics and Astronomy, Northwestern University, 
Evanston, IL 60208}
\author{Loyal Durand}
\affiliation{Department of Physics, University of Wisconsin, Madison, WI 53706}
\author{Douglas W. McKay}
\affiliation{Department of Physics and Astronomy, University of Kansas, Lawrence, KS 66045} 
\date{\today}

\begin{abstract}
We recently derived an explicit expression for the gluon distribution function $G(x,Q^2)=xg(x,Q^2)$ in terms of  the proton structure function $F_2^{\gamma p}(x,Q^2)$  in  leading-order (LO) QCD
 by solving the LO  DGLAP  equation for the $Q^2$ evolution of  $F_2^{\gamma p}(x,Q^2)$ analytically, using a differential-equation method.
We showed that  accurate experimental knowledge of $F_2^{\gamma p}(x,Q^2)$ in a region of Bjorken $x$ and virtuality $Q^2$ is all that is needed to determine the gluon distribution in that region.  We re-derive and extend the results here using a Laplace-transform technique, and show that the singlet quark structure function $F_S(x,Q^2)$ can be determined directly in terms of $G$ from the DGLAP gluon evolution equation.  To illustrate the method and check the consistency of existing LO quark and gluon distributions, we used the published values of  the LO  quark distributions from the CTEQ5L and MRST2001LO  analyses to form $F_2^{\gamma p}(x,Q^2)$, and then solved analytically for $G(x,Q^2)$.  We find that the analytic and  fitted  gluon  distributions from MRST2001LO agree well with each other for all x and $Q^2$, while those from
CTEQ5L differ significantly from each other for large $x$ values, $x\ga 0.03-0.05$, at all $Q^2$. 
We conclude that the published CTEQ5L  distributions are incompatible in this region.  Using a non-singlet  evolution equation, we obtain a sensitive test of quark distributions which holds in both LO and NLO perturbative QCD.  We find in either case that the CTEQ5 quark distributions satisfy the tests numerically for small $x$, but  fail the tests for $x\ga 0.03-0.05$---their use could potentially lead to significant shifts in predictions of quantities sensitive to large $x$. We encountered no problems with the MRST2001LO distributions or later CTEQ distributions.  We suggest caution in the use of the CTEQ5 distributions.  

\end{abstract}


\maketitle


\section{Introduction}\label{sec:introduction}

In a recent paper \cite{bdm1}, we derived an explicit expression for the gluon distribution function $G(x,Q^2)=xg(x,Q^2)$ in the proton in terms of  the proton structure function $F_2^{\gamma p}(x,Q^2)$ for $\gamma^*p$ scattering.  The result was obtained in  leading-order (LO) QCD
 by solving the  Dokshitzer-Gribov-Lipatov-Altarelli-Parisi (DGLAP)  equation \cite{dglap} for the $Q^2$ evolution of  $F_2^{\gamma p}(x,Q^2)$ analytically,  assuming  massless quarks and using a differential-equation method.   The method is model-independent and does not require individual quark distributions, so accurate experimental knowledge of $F_2^{\gamma p}(x,Q^2)$ in a region of Bjorken $x$ and virtuality $Q^2$ is all that is needed to determine the gluon distribution in that region. 
 
 In the present paper, we extend the method using a Laplace-transform technique. We first re-derive the solution for $G(x,Q^2)$ in terms of 
 \be
  F_2^{\gamma p}(x,Q^2)=\sum_{i=1}^{n_f}e_i^2x\left[q_i(x,Q^2)+\bar{q}_i(x,Q^2)\right] \label{F2pdef}
\ee
 using our Laplace transform method.  The same procedure can be used to derive $G(x,Q^2)$ from  the singlet (S) quark distribution function 
 \be
 F_S(x,Q^2)=\sum_{i=1}^{n_f}x\left[q_i(x,Q^2)+\bar q_i(x,Q^2)\right] \label{FS}
 \ee 
 and other combinations of the quark distributions. We then obtain an analytic solution for $F_S(x,Q^2)$
 in terms of $G(x,Q^2)$ by solving the  LO gluon evolution equation exactly.  These are the principal theoretical results in this paper.

To illustrate the method, we have  used the LO CTEQ5L \cite{CTEQ5} 
and MRST2001 \cite{MRST2001} quark distributions to construct the proton $\gamma^* p$ and singlet structure functions as functions of $x$ and $Q^2$, solved analytically for the  corresponding gluon distributions, and compared the results for $G(x,Q^2)$ with those  given in the  fits. We find that the analytic and published distributions agree to high accuracy for MRST2001 LO, but, to our surprise, that they are incompatible for CTEQ5L, differing significantly at large $x$ for all $Q^2$.  We find the same situation when we use the evolution equations for the singlet quark distribution or for individual quarks to solve for $G(x,Q^2)$.  We have not encountered these difficulties in somewhat less-extensive calculations using the  more recent CTEQ6L \cite{CTEQ6} distributions, finding good agreement between the analytic and published gluon distributions. 

To investigate the inconsistencies further, we have used the DGLAP evolution equation for the non-singlet (NS) distributions
\ba
F_{NS}&\equiv& \sum_{i=1}^{n_f}x\left[ u_i(x,Q^2)+\bar u_i(x,Q^2) \right. \nonumber \\
&& \qquad \left. - d_i(x,Q^2)-\bar d_i(x,Q^2)\right], \label{FNS}
\ea
where the sum is over the up-  and down-type quarks and antiquarks in each generation with both types active.
As will be shown, $F_{NS}$ can be used in both leading order (LO) and next-to-leading order (NLO)  to construct  sensitive tests of the quark distributions. 

The tests based on $F_{NS}$  are  satisfied for the MRST2001 LO \cite{MRST2001} and CTEQ6L \cite{CTEQ6} quark distributions which also satisfied the gluon tests above  to high accuracy. However, we find that the LO CTEQ5L and the NLO CTEQ5M quark distributions in the $\overline{\rm MS}$ renormalization scheme fail to satisfy the  NS evolution equation in the same regions at  large $x$ where the gluons gave difficulty. This shows clearly that these published quark distributions are inconsistent at large $x$. 

The errors  in the quark distributions and their incompatibilities with $G$ could well affect predictions for  experimental quantities that are sensitive to large $x$.  We conclude that the CTEQ5L distributions, though very convenient because of the analytic parametrizations given \cite{CTEQ5}, should not be used to make precise predictions at large $x$.


\section{ Preliminaries}\label{sec:preliminaries}

We first introduce  a quantity ${\cal F}_2(x,Q^2)$, which depends {\em only} on the proton structure function $F_2^{\gamma p}(x,Q^2)$, by
\ba
{\cal F}_2(x,Q^2)\!&\equiv&\!\frac{\partial F_2^{\gamma p}(x,Q^2)}{\partial \ln (Q^2)}\nonumber\\
&&\hspace*{-0.5cm}-\frac{\alpha_s}{4\pi}\left\{ 4{F_2^{\gamma p}(x,Q^2)}\!+\!{16\over 3}\left[{F_2^{\gamma p}(x,Q^2)}\ln\frac{1-x}{x}\right.\right.\nonumber\\
&&\hspace*{-0.5cm}\left.\left.
+x\int_x^1\left(\frac{F_2^{\gamma p}(z,Q^2)}{z}-\frac{F_2^{\gamma p}(x,Q^2)}{x}\right){dz\over z-x}\right]\right.\nonumber\\
&&\hspace*{-0.5cm}\left.-\frac{8}{3}x\int_x^1F_2^{\gamma p}(z,Q^2)\left(1+\frac{x}{z}\right)\frac{\,dz}{z^2}\right\}.
\label{AP200}
\ea
We define the corresponding quantities ${\cal F}_S(x,Q^2)$ and ${\cal F}_{NS}(x,Q^2)$ for the singlet and non-singlet quark distributions  obtained by replacing $F_2^{\gamma p}$ by $F_S$ or $F_{NS}$.

The LO DGLAP evolution equations  \cite{dglap} for the  proton structure functions $F_2^{\gamma p}(x,Q^2)$, $F_S(x,Q^2)$, and $F_{NS}(x,Q^2)$ for massless quarks
 can  be written compactly in $x$ space as 
\ba
{\cal F}_2(x,Q^2)\!&=&\! \asoverpi \sum_i e_i^2 x\int_x^1G(z,Q^2)K_{qg}\left({x\over z}\right)\frac{\,dz}{z^2},\quad \label {LOAP2} \\
{\cal F}_S(x,Q^2)\!&=&\! 2n_f\asoverpi x\int_x^1G(z,Q^2)K_{qg}\left({x\over z}\right)\frac{\,dz}{z^2},\label {LOFS}\\
{\cal F}_{NS}(x,Q^2)\!&\equiv&\! 0,\label {LOFNS}
\ea
where the sum over quark charges in Eq.\ (\ref{LOAP2}) includes both quarks and antiquarks, and $n_f$  in Eq.\ (\ref{LOFS}) is the number of active quarks. ${\cal F}_{NS}$ vanishes only if both members of a quark family are massless or can otherwise be treated as active.  The LO $g\rightarrow q$ splitting function is given by
\ba
K_{qg}(x)=1-2x+2x^2. \label{splittingfunction}
\ea
Since
\be
F_2^{\gamma p}(x,Q^2)=\frac{5}{18}F_S(x,Q^2)+\frac{1}{6}F_{NS}(x,Q^2),
\ee
we should be able to derive $G(x,Q^2)$ consistently in LO from either of Eqs.\ (\ref{LOAP2}) or (\ref{LOFS}) provided Eq.\ (\ref{LOFNS}) is satisfied. 

We note that Eqs.\ (\ref{LOFS}) and (\ref{LOFNS}) continue to hold at next-to-leading order with modified quark and gluon splitting functions in Eqs.\ (\ref{AP200}) and (\ref{LOFS}). These are given  in the $\overline {\rm MS}$ scheme in Eqs.\ (4.7a), (4.7b) and (4.7c) of Floratos {\em et al.} \cite{Floratos}.

To simplify the equations, we rewrite \eq{LOAP2} as
\ba
{\calF \cal F}(x,Q^2)=x\int_x^1G(z,Q^2)K_{qg}\left({x\over z}\right)\frac{\,dz}{z^2}\,\label{finalG},
\ea
where we have introduced ${\cal F\calF}(x,Q^2)$, which is defined as
\ba
{\calF \cal F}(x,Q^2)&=& \left(\asoverpi \sum_ie_i^2\right)^{-1}
{\cal F}_2(x,Q^2)\label{newcalF}.
\ea
Similarly, we define ${\calF \cal F}_S(x,Q^2)$ in terms of the singlet quark distribution $F_S(x,Q^2)$
 by
\ba
{\calF \cal F_S}(x,Q^2)&=& \left(2n_f \asoverpi \right)^{-1}
{\cal F}_2(x,Q^2), \label{FF_S}
\ea
and analogous quantities with appropriate prefactors for other structure functions such as $F_{2 (3)}^{\gamma Z}$, $F_{2 (3)}^Z$, and $F_{2 (3)}^{W^\mp}$.

We note also that if we wish to evolve a single massless quark distribution in LO,   we can simply replace $F_2^{\gamma p}$ in Eq.\ (\ref{AP200}) by $Q_i=xq_i(x,Q^2)$ and ${\cal F}{\cal F}$ in \eq{newcalF} by the appropriate
\ba
{\calF \cal F}_i(x,Q^2)&=& \left(\asoverpi\right)^{-1}
{\cal F}_i(x,Q^2)\label{newcalFi},
\ea
so that   
\ba
{\calF \cal F}_i(x,Q^2)=x\int_x^1G(z,Q^2)K_{qg}\left({x\over z}\right)\frac{\,dz}{z^2}\,\label{finalGi},
\ea
with $i=u,\bar u, d,\bar d,c,\bar c, s,\bar s$.

Since the gluon content of \eq{finalGi} is independent of the type of quark, \eq{finalGi} has the 
 important consequence that, in LO,
\be
{{\calF \cal F}_i(x,Q^2)\over {\calF \cal F}_j(x,Q^2)}=1,\label{rtest}
\ee
for all  $x$ at a fixed virtuality $Q^2$,  and  any choice of $i$ and $j$. As we will soon see, this turns out to be a very sensitive test of compatibility of quark distributions with the LO QCD evolution equations.  

An alternate form of this relation, which is readily extendable to NLO, follows from the evolution equation for the non-singlet distribution $F_{\rm NS}(x,Q^2)$, ${\cal F}{\cal F}_{\rm NS}(x,Q^2)=0$. Splitting the sums in Eq.\ (\ref{FNS})
into sums over up- and down-type quarks,  we find that the vanishing of  ${\cal F}{\cal F}_{\rm NS}(x,Q^2)$ implies that
\be
{{\calF \cal F}_{\rm dtype}(x,Q^2)\over {\calF \cal F}_{\rm utype}(x,Q^2)}=1\label{NLO}
\ee
for {\em all } $x$ at any given virtuality $Q^2$. This furnishes us with an exact numerical test of NLO perturbative QCD. We will later see that, while the MRST2001 and CTEQ6   parton distributions satisfy our constraints for all $Q^2$ and x, the LO CTEQ5L and NLO  $\overline {\rm MS}$ CTEQ5M quark distributions  satisfy these tests for small $x$, but fail badly at large $x$.  


\section{Solution of the  LO evolution equation for the proton structure function $F_2^{\gamma p}(x,Q^2)$}\label{sec:DGLAP}

We now re-derive our analytic solution \cite{bdm1} of \eq{finalG} for the LO gluon distribution, $G(x,Q^2)$ using a new Laplace-transform method. Introducing the coordinate transformation 
\be v\equiv \ln(1/x),
\ee
we define functions $\hat{G}$, $\hat{K}_{qg}$, and $\hat{\cal F}$ in $v$-space by
\ba
\hat G(v,Q^2)&\equiv& G(e^{-v},Q^2)\nonumber\\
\hat K_{qg}(v)&\equiv& K_{qg}(e^{-v})\nonumber\\
\hat \calF (v,Q^2)&\equiv& {\cal F\calF}(e^{-v},Q^2).\label{hats}
\ea
Explicitly, from \eq{splittingfunction}, we see that
\ba
\hat K_{qg}(v)=1-2e^{-v}+2e^{-2v}\label{Khat}. 
\ea
Further,  we can write
\ba
\hatF\!&=&\!\int^v_0\hat G(w,Q^2)e^{-(v-w)}\hat K_{qg}(v-w)\,dw\nonumber\\
\!&=&\! \int^v_0\hat G(w,Q^2)\hat H(v-w)\, dw,\label{DGLAP1}
\ea
where
\ba
\hat H(v)&\equiv&e^{-v}\hat K_{qg}(v)\nonumber\\
&=& e^{-v}-2e^{-2v}+2e^{-3v}.\label{H}
\ea

We introduce the notation that the  Laplace transform of a function  $\hat H(v)$ is given by $h(s)$, where 
\ba
h(s)\equiv{\cal L}[\hat H(v);s]= \int_0^\infty \hat H(v)e^{-sv} dv,
\ea
with the condition $ \hat H(v)=0 \quad \rm{for \ } v<0$. 
The convolution theorem for Laplace transforms relates the transform of a convolution of functions $\hat G$ and $\hat H$ to the product of their transforms $g(s)$ and $h(s)$, so that 
\be
{\cal L}\left[\int_0^v \hat G(w)\hat H(v-w)\,dw;s\right]=g(s)\times h(s), \label{direct_convolution}
\ee
or conversely, the inverse transform of a product to the convolution of the original functions, giving  
\ba
{\cal L}^{-1}[g(s)\times h(s);v]&=& \int_0^v \hat G(w)\hat H(v-w)\,dw\nonumber\\
& = &\int_0^v \hat H(w)\hat G(v-w)\,dw. \label{convolution}
\ea

Equation (\ref{direct_convolution})  allows us to write the Laplace transform of  \eq{finalG}
as
\ba
f(s,Q^2)&=&g(s,Q^2)\times h(s),\label{LT1}
\ea
where
\be
f(s,Q^2)={\cal L}[\hat{\cal F}(v,Q^2);s] \label{f(s)}
\ee
and $g(s,Q^2)$ and $h(s)$ are the transforms of the functions $\hat{G}$ and $\hat{H}$ on the right-hand-side of  Eq.\ (\ref{DGLAP1}). Solving \eq{LT1} for $g$, we find that 
\be
g(s,Q^2)=f(s,Q^2)/h(s). \label{g_solution}
\ee

We will generally not be able to calculate the inverse transform of $g(s,Q^2)$ explicitly, if only because $f(s,Q^2)$ is determined by a numerical integral of the experimentally-determined function $\hat{\cal F}(v,Q^2)$. However, if we regard $g(s,Q^2)$ as the product of the two functions $f(s,Q^2)$and $h^{-1}(s)$ and take the inverse Laplace transform using the convolution theorem in the form in Eq.\ (\ref{convolution}) and the {\em known inverse} ${\cal L}^{-1}[f(s,Q^2);v]={\hat \calF}(v,Q^2)$, we find that
\ba
{\hat G}(v,Q^2)&=&{\cal L}^{-1}[f(s,Q^2)\times h^{-1}(s);v] \nonumber \\
&=&\int_0^v{\hat \calF}(w,Q^2)\hat J(v-w)\,dw
\ea
where $\hat {J}(v)$ is a new auxiliary function, defined by 
\be
\hat J(v) \equiv {\cal L}^{-1}[h^{-1}(s);v]. \label{Htilde}
\ee

The calculation of $h(s)$ and the inverse Laplace transform of of $h^{-1}(s)$ are straightforward, and we find that
\ba
\hat J(v)&=&3\delta(v)+ \delta'(v)\nonumber\\
&&-e^{-3 v/2}\left({6\over \sqrt 7} \sin \left[ {\sqrt 7\over 2}v\right]\nonumber\right.\\
&&\qquad\qquad\left.+2\cos\left[{\sqrt 7\over 2}v\right] \right).\label{Hminus1}
\ea
We  therefore obtain an explicit  solution for the gluon distribution $G(v,Q^2)$ in terms of the integral 
\ba
\hat G(v,Q^2)&=&\int_0^v {\hat \calF}(w,Q^2)\hat J(v-w)\,dw\nonumber\\
&= &3{\hat \calF}(v,Q^2)+{\partial {\hat \calF}(v,Q^2)\over \partial v }\nonumber\\
&&\quad-\int_0^v {\hat\calF}(w,Q^2) e^{-3 (v-w)/2}\times\nonumber\\
&&\qquad\left({6\over \sqrt 7} \sin \left[ {\sqrt 7\over 2} (v-w)\right]\right.\nonumber\\
&&\qquad\quad\left.+2\cos\left[{\sqrt 7\over 2}(v-w)\right]\! \right)dw.\label{Gofv}
\ea

Transforming \eq{Gofv} back into $x$ space, we find that the gluon probability distribution $G(x,Q^2)$ is given in terms of the  function $\cal{F}\cal{F}$, assumed to be known, by
\ba
G(x,Q^2)&=&\!3\,{ \calF}{ \calF}(x,Q^2)-x{\partial { \calF}{ \calF}(x,Q^2)\over \partial x }\nonumber\\
&&\quad-\int_x^1 {\calF}{ \calF}(z,Q^2) \left({x\over z}\right)^{3/2}\times\nonumber\\
&&\qquad\left\{{6\over \sqrt 7} \sin \left[ {\sqrt 7\over 2} \ln{z\over x}\right]\right.\nonumber\\
&&\quad\qquad\left.+2\cos\left[{\sqrt 7\over 2}\ln{z\over x}\right] \right\}{dz\over z}.\label{Gofx}
\ea
It can be shown without difficulty that this expression for $G$ is equivalent to that which we derived in \cite{bdm1}, where we first converted the evolution equation for $F_2^{\gamma p}(x,Q^2)$ into a differential equation in $v$, and then solved that equation explicitly. 

Even if we cannot  integrate the last term of \eq{Gofx} analytically, the usual case,  it can be  integrated numerically, so that we {\em always} have an explicit  solution which can be evaluated to the numerical accuracy to which $ { \calF}{ \calF}(x,Q^2)$ is known.  We emphasize that this solution for $G$ is derived from $F_2^{\gamma p}$ on the assumption that the quarks are either massless, or effectively so, a situation that holds in general for $Q^2\gg 4M_i^2$, where mass effects due to the $i^{\rm th}$ (massive) quark are negligible, or for a treatment of mass effects such as that used in CTEQ5 \cite{CTEQ5} as discussed in Sec.\ \ref{sec:compatibility}.
\footnote{We will discuss mass effects in detail elsewhere.} 

Leading-order solutions for $G(x,Q^2)$ of exactly the same form follow from the evolution equations for the singlet structure function $F_S$ and the individual quark distribution functions, with $ { \calF}{ \calF}(x,Q^2)$ replaced in Eq.\ (\ref{Gofx}) by $ { \calF}{ \calF}_S(x,Q^2)$ or $ { \calF}{ \calF}_i(x,Q^2)$, as noted earlier. The latter can be used to obtain expressions for $G(x,Q^2)$ in terms of the structure functions for weak scattering processes, allowing the incorporation of other data sets.


\section{ Solution of the  LO  gluon evolution equation}\label{sec:gluon_equation}

To obtain an expression which relates the singlet structure function $F_S(x,Q^2)$ indirectly to  experiment, we use the  DGLAP equation for the $Q^2$ evolution of the gluon distribution $G(x,Q^2)$. This relates $F_S$ to $G$ which is determined by $F_2^{\gamma p}$ as shown in the preceding section.
Again, we simplify the notation by writing this equation in terms of the (rescaled) $q\rightarrow g$ splitting function  $K_{gq}(x)$ and a quantity  ${\cal G\cal G}(x,Q^2)$ as
 \ba
{\cal G\cal G}(x,Q^2)&=&x\int _x^1F_S(x,Q^2)K_{gq}(x/z)\,{dz\over z^2}.
\ea
Here
\be
K_{gq}(x)\equiv{2\over x }-2+x,
\ee
and
\ba
{\cal G\cal G}(x,Q^2)\!&\equiv&\!{3\over 8}\left\{\left(\asoverpi \right)^{-1}{\partial G(x,Q^2)\over \partial \ln Q^2}\right.\nonumber\\
&-&\left.\! \!G(x,Q^2)\left({1\over 3}(33-2n_f)+12\ln(1-x)\right)\right.\nonumber\\
&-&\left.\!\!12 x\!\left[\int_x^1 G(z,Q^2)\left( {z\over x}-2+{x \over z}-  \left({x\over z}\right)^2\right)\!{dz\over z^2}      \right.\right.\nonumber\\
&&\hspace*{-1cm}\left. \left.+\int_x^1\!\left(G(z,Q^2)\!-\!G(x,Q^2)  \right)\!\left( {z\over z-x}\right)  {dz\over z^2}\right]\!\right\},\label{calG}
\ea
 for $n_f$ effectively massless active quark flavors. 
As before, going to $v$ space, we define the quantities 
\ba
\hat F_S(v,Q^2)&\equiv& F_S(e^{-v},Q^2), \nonumber\\
{\hat{\cal G}}(v,Q^2)&\equiv& {\cal G\cal G}(e^{-v},Q^2), \nonumber\\
\hat H_{gq}(v,Q^2)&\equiv& e^{-v}K_{gg}(e^{-v}) \nonumber\\
&=&e^{-v}\left(2e^v-2+e^{-v}\right) \nonumber\\
&=&2-2e^{-v}+e^{-2v}.
\ea
The evolution equation then involves a convolution of $\hat{ F}_S$ and $\hat{ H}_{gq}$.

Proceeding as  before, we take the Laplace transform of both sides, solve for the transform $f_S$ of $\hat{F}_S$, invert the transform using the convolution theorem again, and find that
\ba
\hat F_S(v,Q^2)&=& {\partial {\hat{\cal G}}(v,Q^2)\over \partial v}+\int_0^v{\hat{\cal G}}(w,Q^2)e^{-3(v-w)/2}\times\nonumber\\
&&\left\{ {6\over \sqrt 7}\sin\left[ {\sqrt 7\over 2} (v-w) \right]\right.\nonumber\\
&&\left.-2\cos\left[ {\sqrt 7\over 2} (v-w) \right]\right\}\,dw.
\ea
Transforming back into $x$ space, $F_S(x,Q^2)$ is given by
\ba
F_S(x,Q^2)&=&- x{\partial { \cal G}{ \cal G}(x,Q^2)\over \partial x }\nonumber\\
&&\quad+\int_x^1 {\cal G}{ \cal G}(z,Q^2) \left({x\over z}\right)^{3/2}\times\nonumber\\
&&\qquad\left\{{6\over \sqrt 7} \sin \left[ {\sqrt 7\over 2} \ln{z\over x}\right]\right.\nonumber\\
&&\quad\qquad\left.-2\cos\left[{\sqrt 7\over 2}\ln{z\over x}\right] \right\}{dz\over z}.\label{FSofx}
\ea

${{\cal G}{\cal G}}(x,Q^2)$ is a known function of the gluon distribution, $G(x,Q^2)$, which, in turn, is a known function of the proton structure function $F_2^{\gamma p}(x,Q^2)$ that is measured in deep inelastic scattering. $F_S(x,Q^2)$ is therefore related back to $F_2^{\gamma p}(x,Q^2)$ through $G(x,Q^2)$ and the analysis in the preceding section, and could be used, in turn, to determine $G(x,Q^2)$ through Eq.\ (\ref{Gofx}) with ${\cal F}{\cal F}_S$ replacing ${\cal F}{\cal F}$. 

It is important to recognize that  our method is essentially model-independent.   When different parametrizations in $x$ and $Q^2$ are used to fit experimental $F_2^{\gamma p}$ data, the LO gluon and $F_S$ distributions derived from the different fits must  agree with each other to  the accuracy of the fits in the regions in which the data exist.  The results can only depend on the functional forms used to fit $F_2^{\gamma p}(x,Q^2)$  when they are extrapolated  to $x$ and $Q^2$ outside that region. 


\section{Numerical illustration of the method, and tests of LO gluon and quark distributions  }\label{sec:compatibility}

To test  of our methods, we have applied them to published parton distributions, using the quark distributions to construct $F_2^{\gamma p}$ and $F_S$ and then solving  \eq{Gofx} to find $G$. The methods work well, and reproduce the published gluons distributions except in the case of  CTEQ5 \cite{CTEQ5}. In that case, we found to our surprise that there are problems at high $x$ with the LO CTEQ5L  distributions.  While these are superseded by more recent CTEQ distributions \cite{CTEQ6,cteq6.5}, they seem still to be used in some calculations, most likely because they have been parametrized analytically in a form convenient for calculation. 
Similar tests applied to the  MRST2001 \cite{MRST2001} and the CTEQ6L \cite{CTEQ6} LO distributions uncovered no problems.   

The results illustrate the sensitivity of the analytic methods in testing parton distributions, and may be of wider use for that purpose, as well as for deriving $G$ directly from data as originally proposed in \cite{bdm1}.  We present them here as a demonstration, and as a caution.

We begin by illustrating the use of the analytic expression in  \eq{Gofx} to derive $G(x,Q^2)$ from $F_2^{\gamma p}(x,Q^2)$ in the case of CTEQ5L \cite{CTEQ5}. We take the  published LO CTEQ5L  quark distributions as our basic input, and use these distributions to calculate the proton structure function
$F_2^{\gamma p}(x,Q^2)=\sum_ie_i^2x[q_i(x,Q^2)+\bar{q}_i(x,Q^2)]\label{F2pdef2}$  needed in 
${\cal F}{\cal F}(x,Q^2)$ in \eq{newcalF}.  Next, we solve \eq{Gofx} for the $G$ generated by this $F_2^{\gamma p}$, labelling the solution $G_{\rm Analytic}(x,Q^2)$, and compare the results with the published gluon distributions.

We  follow the procedures of the CTEQ5 group in these calculations. The parton splitting functions used are those for massless quarks.  Quark mass effects are included only in an approximate, $x$-independent way, with $xq_i$ and $x\bar{q}_i$ taken as zero for a massive quark $i$ when $Q^2<M_i^2$, and the quark treated as fully  active when $Q^2>M_i^2$, with $q_i$ and $\bar{q}_i$ included in the calculations of the structure functions $F_2^{\gamma p}$ and $F_S$. The charge factors $\sum_ie_i^2$ in Eqs.\, (\ref{LOAP2}) and  (\ref{newcalF}), and the numbers of active quarks $n_f$ in Eqs.\ (\ref{LOFS}), (\ref{FF_S}), and later in Eq.\ (\ref{GNS}), then change discontinuously at each threshold, but remain constant between thresholds. Despite these discontinuities, the quark and gluon distributions are continuous.
\footnote{ The quark and gluon  distributions are continuous solutions of the evolution equations, which are first-order differential equations in $\ln Q^2$ and integral equations in $x$. The discontinuities on the right-had sides of Eqs.\ (\ref{LOAP2}) and (\ref{LOFS}) as $n_f$ changes at a threshold are reflected 
in the final results  by discontinuous changes in the derivatives of the structure functions $\partial F/\partial\ln Q^2$ on the left-hand sides of these equations. In particular, the heavy quark distributions $xq_i(x,Q^2)$ and $x\bar{q}_i(x,Q^2)$ are identically zero for $Q^2<M_i^2$, then  initially rise linearly from zero with increasing $\ln Q^2$ for $Q^2\ge M_i^2$. The contributions of $\partial (xq_i)_i/\partial\ln Q^2$ and $\partial (x\bar{q}_i)_i/\partial\ln Q^2$ to  $\partial F/\partial\ln Q^2$, identically zero for $Q^2<M_i^2$, therefore jump to  non-zero values for $Q^2\ge M_i^2$.} 

Because the mass effects do not depend on $x$ in this approach, the evolution equations retain the massless  form between thresholds,  and we can simply use the results derived above for massless quarks in our analysis, treating the distinct inter-threshold regions in $Q^2$ separately.
\footnote{Mass effects are treated the same way by CTEQ6 \cite{CTEQ6}. The more accurate treatments of heavy-quark masses in MRST2001 \cite{MRST2001} and the recent CTEQ analyses beginning with CTEQ6.5 \cite{cteq6.5} involve $x$-dependent effects. These will not affect the tests of the MRST2001 distributions given later using only the massless quarks.}

In these calculations, we use the LO form of $\alpha_s(Q^2)$ used in CTEQ5L \cite{alphacteq}, 
\ba
\alpha_s(Q^2)
&=&\frac{4 \pi}{\beta_{0}\ln(Q^2/\Lambda^2)},
\label{alphaLO}\\
\beta_{0}&=& 11-{2\over3}n_f,
\ea
with $n_f=5$ and $\Lambda_5=146$ MeV for $Q>4.5$ GeV, $n_f=4$ and $\Lambda_4=192$ MeV for 1.3 GeV $<Q\leq 4.5$ GeV, and $n_f=3$ and $\Lambda_3=221$ MeV for $Q<1.3$ GeV. These parameters give a continuous $\alpha_s(Q^2)$ with $\alpha_s(M_Z^2)=0.127$ \cite{CTEQ5}. 

The same methods can be used in LO to determine the gluon distributions $G_i$ generated by individual quark distributions, or by other combinations of the quarks, in particular, by combinations of the massless quarks $u,\,d,\,s$ and by the non-singlet distributions, as will be seen below.

Following these procedures, the gluon distribution $G_{\rm Analytic}(x,Q^2)$ we  derive analytically from the calculated structure functions  $F_2^{\gamma p}$ or  $F_S$, or  from other quark combinations,  should agree  to the accuracy of the calculations with $G_{\rm CTEQ5L}(x,Q^2)$ at all $x$ and $Q^2$.

\begin{figure}[h,t,b] 
\begin{center}
\mbox{\epsfig{file=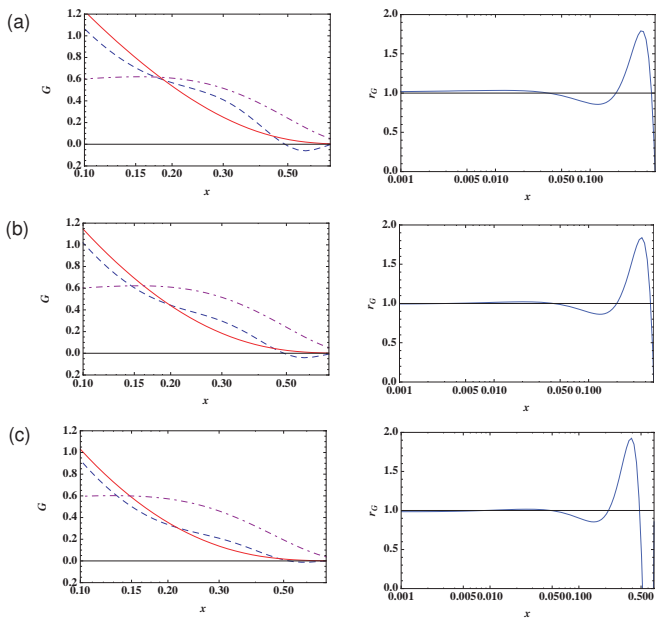,width=3.4in%
}}
\end{center}
\caption[]{Left -hand column: comparison of the $G_{\rm Analytic}$ (blue dashed curves) obtained by solving  \eq{Gofx} using the CTEQ5L $F_2^{\gamma p}(x,Q^2)$ as input, with  $G_{\rm CTEQ5L}$ (red curves). The $u$-quark distributions $U(x,Q^2)=xu(x,Q^2)$ (purple dot-dashed curves) are shown to give an independent scale at the same $x$ and $Q^2$.  (a) $Q^2= 5$ GeV$^2$, $n_f=4$. (b) $Q^2=20$ GeV$^2$, $n_f=4$.  (c)  $Q^2= 100$ GeV$^2$, $n_f=5$. Right-hand column: the ratio $r_G=G_{\rm Analytic}/G_{\rm CTEQ5L}$ at the same $Q^2$ as (a), (b), (c). The ratios should equal 1 for all $x$ and $Q^2$.  } \label{GGcomp}
\end{figure}

The results of the calculations based on $F_2^{\gamma p}$ are shown in Fig.\  \ref{GGcomp}. The left-hand column compares $G_{\rm Analytic}$ (blue dashed curves) with $G_{\rm CTEQ5L}$ (red solid curves) at $Q^2=5$, 20, and 100 GeV$^2$,  (a), (b), and (c), respectively.  All of the active, physically-relevant quarks are included in the input at each value of $Q^2$, with $n_f=4$ for (a) and (b) with $u,\,d,\,s,\,c$ active, and $n_f=5$ in (c) with $b$ now also  active. For a scale comparison, we include the CTEQ5L up-quark distribution, $U(x,Q^2)=xu(x,Q^2)$.

The analytic and fitted $G$\,s do not  agree well at large $x$,  with significant differences between them on the scale of $G$ and of the $u$-quark distribution  for $0.05 \lesssim  x \lesssim 0.6$.  We do not believe that the calculations are reliable beyond $x\approx 0.6$: $G$ is very small , but the input $u$ and $d$ distributions and the individual integrals in $\cal{F}\cal{F}$ are not, and the calculation in Eq.\ (\ref{Gofx}) involves large cancellations and becomes  sensitive to small errors in the inputs.  

The same discrepancies between the two $G$\,s are illustrated  differently in the right-hand column of Fig.\ \ref{GGcomp} where we plot the ratio
\be
 r_{G}={G_{\rm Analytic}(x,Q^2)\big / G_{\rm CTEQ5L}(x,Q^2)}. \label{Gratio}
\ee
In all cases, we find  that $r_G$ is essentially constant and equal to one  for all  $x\la 0.01$ for small virtuality, and for all  $x\la 0.05$ for large virtuality, showing good agreement between analytic and fitted gluon distributions at small $x$.  There are again significant deviations from one at larger $x$, contrary to theoretical requirements.  

We emphasize that the reason we find {\em negative\/} values of $G_{\rm Analytic}$ or $r_G$ for some $x\greaterabout 0.5$ is {\em not\/} that $F_2^{\gamma p}(x,Q^2)$ goes negative, but rather that ${\cal F}{\cal F}(x,Q^2)$ goes negative. As can be seen from \eq{AP200}, this is a sensitive function of the {\em difference} between $\partial F_2^{\gamma p}(x,Q^2)/ \partial \ln (Q^2)$ and the LO convolution integral of  $F_2^{\gamma p}(x,Q^2)$.  In spite of the fact that the CTEQ5L quark distributions that go into  $F_2^{\gamma p}(x,Q^2)$ are all  positive, the resulting combination ${\cal F}{\cal F}(x,Q^2)$ becomes negative for large $x$.  The appearance of negative values for the gluon distribution function suggests strongly that the differences between $G_{\rm Analytic}$ and $G_{\rm CTEQ5L}$ in Fig.\ \ref{GGcomp} result from problems with the input quark distributions, hence with $G_{\rm Analytic}$, and are less likely to arise directly from problems with $G_{\rm CTEQ5L}$. 

We find very similar deviations  if we use the singlet structure function $F_S(x,Q^2)$ in Eq.\ (\ref{FS}) as input, and then solve for $G(x,Q^2)$ using  Eq.\ (\ref{Gofx}) with ${\cal F}{\cal F}_S$ replacing ${\cal F}{\cal F}$.

The discrepancies between the analytic and fitted $G$\,s are similar in magnitude to the differences between the CTEQ5 and more recent CTEQ and MRST gluon distributions in this region, and to the changes in the distributions that resulted from the addition of inclusive jet data to those analyses. Changes of this size are clearly significant.

\begin{figure}[h,t,b] 
\begin{center}
\mbox{\epsfig{file=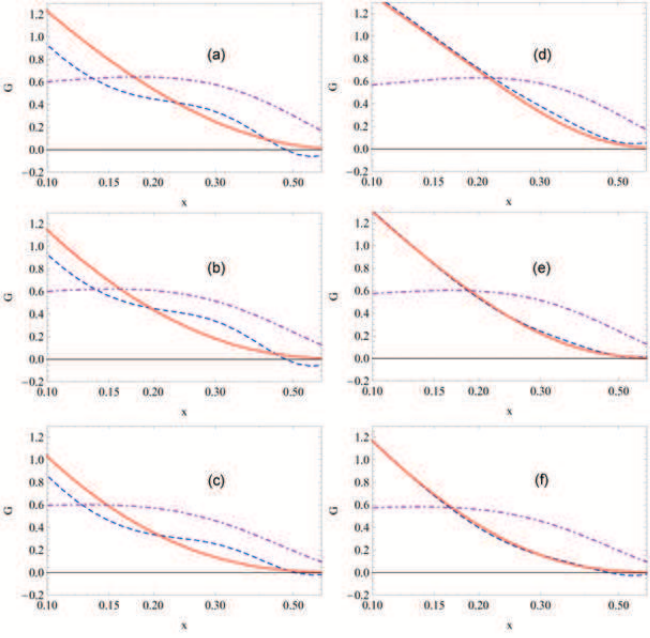,width=3.4in%
,bbllx=0pt,bblly=0pt,bburx=610pt,bbury=620pt,clip=%
}}
\end{center}
\caption[]{Plots of the $n_f=2$ $G_{\rm ud}(x,Q^2)$ (blue dashed curves), $G_{\rm fitted}(x,Q^2)$ (red curves), and the $u$-quark distribution $U(x,Q^2)=xu(x,Q^2)$ (purple dot dashed curves) vs. $x$, for   $x>0.1$. Left-hand column: $G_{\rm fitted} = G_{\rm CTEQ5L}$ distributions for (a) $Q^2=5$ GeV$^2$,  (b) $Q^2=20$ GeV$^2$, (c)  $Q^2=100$ GeV$^2$. Right-hand column: $G_{\rm fitted} = G_{\rm MRST2001}$ LO distributions for (d) $Q^2=5$ GeV$^2$,  (e) $Q^2=20$ GeV$^2$, (f)  $Q^2=100$ GeV$^2$.  } \label{G&U}
\end{figure}

Since the major contributions to $F_2^{\gamma p}(x,Q^2)$  at large $x$ are from the valence quarks $u$ and $d$, we have  done a separate set of calculations using the singlet combination $F_{S,ud}=x(u+\bar{u}+d+\bar{d})$ and Eq.\ (\ref{Gofx}) with ${\cal F}{\cal F}_{S,ud}$ replacing ${\cal F}{\cal F}$. No mass effects need to be included in this case.  The results of the calculations, called $G_{ud}(x,Q^2)$,  are shown in Fig.\ \ref{G&U} for both CTEQ5L and MRST2001 LO. 

Discrepancies between $G_{\rm CTEQ5L }$ and the $G$ analytically reconstructed from $F_{S,ud}$, here labeled $G_{ud}(x,Q^2)$, are again evident in the figure. The discrepancies have the same pattern and are roughly the same sizes as those seen in Fig.\ \ref{GGcomp}, suggesting that the problems originate with the $u$ and $d$ distributions that drive the initial DGLAP evolution at large $x$. The results for MRST2001 LO show agreement between the calculated $G_{ud}$ and $G_{\rm MRST2001}$ at the level of accuracy of the calculation, which required the parametrization of numerical data from \cite{MRSTweb}.

To  exclude the possibility that the problems with CTEQ5 can be eliminated by going to NLO perturbative QCD, we also looked at the evolution of the  non-singlet structure function $F_{NS}(x,Q^2)$, and at the gluon distribution from the evolution of the singlet structure function $F_S(x,Q^2)$,  using the CTEQ5M NLO distributions as input.  We found a similar $x$-dependence for the gluon ratio obtained from the  NLO singlet  distribution, hence, continuing problems.  The NS combination, for both LO and NLO, is discussed in the next section, and shows clearly that there are problems with the CTEQ5L quark distributions. 

We note finally that, to reduce the possibility that the discrepancies that  we have encountered are a artifact of the published parametrization of the CTEQ5L parton distributions \cite{CTEQ5}, we have also compared them to the corresponding numerical CTEQ5L distributions from the Durham web site \cite{MRSTweb} and concluded that they are numerically  compatible. 

Because the $G_{\rm Analytic}(x,Q^2)$ derived from either $F_2^{\gamma p}(x,Q^2)$ or $F_S(x,Q^2)$ fails to match $C_{\rm CTEQ5L}(x,Q^2)$ well at large $x$, we cannot expect the inverse problem, the analytic derivation of $F_S$ from $G$ to work well for the CTEQ5L distributions, and do not include such calculations here.


\section{ Tests of   LO and NLO non-singlet distributions}\label{sec:NLO_tests}

 For massless quarks in LO, the relation in \eq{rtest} holds for any quark or anti-quark pair $i$ and $j$, 
\be
{{\calF \cal F}_i(x,Q^2)\over {\calF \cal F}_j(x,Q^2)}=1.\label{rtest2}
\ee

As discussed earlier, this can be generalized readily to NLO  using the non-singlet evolution equation ${\cal F}_{NS}(x,Q^2)=0$, Eq.\ (\ref{LOFNS}) with  the NLO NS splitting functions of Floratos {\em et al.} \cite{Floratos}. If we separate the sum in Eq.\ (\ref{FNS}) into sums over $u$- and $d$-type quarks, defined as 
 \ba
 \!\!\!\!\!\!\!\!\!u_{\rm type}&\!\!\!\equiv &\!\!\!x\!\left[u(x,Q^2)\!+\bar u(x,Q^2)+\!c(x,Q^2)+\!\bar c(x,Q^2)\right],\label{utype2}\\
\!\!\!\!\!\!\!\!\!d_{\rm type}&\!\!\!\equiv &\!\!\!x\!\left[ d(x,Q^2)\!+\bar d(x,Q^2)+\!s(x,Q^2)+\!\bar s(x,Q^2)\right],\label{dtype2}
\ea
this relation becomes 
\be
{\cal F}{\cal F}_{NS}(x,Q^2)={\cal F}{\cal F}_{\rm utype}(x,Q^2)-{\cal F}{\cal F}_{\rm dtype}(x,Q^2)=0,,
\ee
where the two terms are to be calculated separately using the same NLO splitting functions as for the full $F_{NS}$. Dividing by the $u$-type term, we obtain the ratio 
\be
r_{\rm NS}\equiv 
{{\calF \cal F}_{\rm dtype}(x,Q^2)\over {\calF \cal F}_{\rm utype}(x,Q^2)}=1,\label{NLO1}
\ee
a relation true for both LO and NLO when all the relevant quarks are active.  

To use this relation to check the consistency of the quark distributions, we insert the appropriate LO  or NLO  quark and anti-quark distributions in Eqs.\ (\ref{utype2}) and(\ref{dtype2}) and use them  to generate $r_{\rm NS}(x,Q^2)$ numerically using the LO or NLO splitting functions. 

In the case of the NLO CTEQ5M $\overline{\rm MS}$ distributions, we use 
\ba
\alpha_s(Q^2)
&=&\frac{4 \pi}{\beta_{0}\ln(Q^2/\Lambda^2)}\times\nonumber\\
&&\left[1-{2\beta_1 \over \beta_0^2}\frac{\ln[\ln(Q^2/\Lambda^2)]}{\ln(Q^2/\Lambda^2)}  \right],
\label{alphaNLO}\\
\beta_{0}&=& 11-{2\over3}n_f\\
\beta_{1}&=& 51-{19\over3}n_f,
\ea
with $\alpha_s(M_z^2)=0.118$ for $n_f=5$,  matching to the $\alpha$'s for $n_f=4$ and 3 at $Q=4.5$ and 1.3 GeV. 

\begin{figure}[h,t,b] 
\begin{center}
\mbox{\epsfig{file=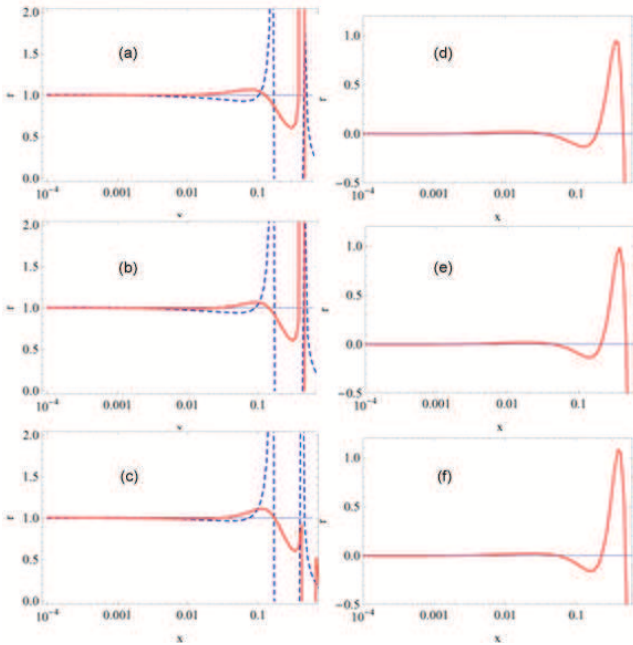,width=3.4in%
,bbllx=0pt,bblly=0pt,bburx=590pt,bbury=630pt,clip=%
}}
\end{center}
\caption[]{ Left-hand column: the LO and NLO non-singlet $d$-type to $u$-type ratio of Eq.\ (\ref{NLO1})   for $n_f=4$ and  $Q^2=5$, 20, and 100 GeV$^2$, (a), (b), and (c), respectively. The ratio should be 1 at all $x$ and $Q^2$.
The solid (red) curves are LO, calculated from  \eq{utype2},  \eq{dtype2}, and \eq{NLO} using  CTEQ5L data for quark distributions . The dashed (blue) curves are NLO calculated for the  $\overline{\rm MS}$ CTEQ5M quark distributions \cite{CTEQ5}.  Right-hand column:
  $r_{NS,G}=G_{NS}(x)/G_{\rm CTEQ5L}(x)$ vs. $x$, for  $Q^2=5$, 20, and 100 GeV$^2$,  (d), (e), and (f), respectively.   The non-singlet ``gluon'' distribution $G_{ NS}(x)$ calculated from $u$ and $d$ quarks   is defined in the text in \eq{GNS} and is  expected to be zero.  \label{fig:doveru}}
\end{figure}

 In the left-hand column of Fig. \ref{fig:doveru}, we plot 
$r_{\rm NS}$ vs. $x$, for  $Q^2=5$, 20, and 100 GeV$^2$ and $n_f=4$. The solid (red) curves were calculated using the LO CTEQ5L quark distributions as input, and the dashed (blue) curves,  the NLO CTEQ5M quark distributions in the $\overline{\rm MS}$ renormalization scheme. We see that the patterns are very similar for all $Q^2$, with ratios $r_{\rm NS}\approx1$ in absolute normalization for all $x\la0.01$, and with very large deviations occurring at large $x$.  This result, independent of the value of $Q^2$, shows explicitly that there are problems with the quark distributions at large $x$: the NS evolution equation is not satisfied in either leading or next-to-leading order.

To look at the size of the deviations in a more quantitative way, we introduce the notion of  a LO ``non-singlet gluon'' distribution $G_{ NS}(x,Q^2)$, defined by the equation 
\ba
\!\!\!\!\!\!{\cal F}_{NS}(x,Q^2)\!\!\!&=&\!\!\!2n_f\asoverpi x\!\int_x^1G_{NS}(z,Q^2)K_{qg}\left({x\over z}\right)\frac{\,dz}{z^2}.\ \label {GNS}
\ea
$G_{NS}=0$ if ${\cal F}_{NS}(x,Q^2)$ is a solution of the non-singlet DGLAP equation for an even number of effectively massless quarks.  Further, $G_{NS}$ is determined mainly by the $u$ and $d$ distributions at large $x$, so measures the accuracy of those distributions.

The normalization on the right-hand side of \eq{GNS} has been chosen so  that this equation is the formal counterpart  of the singlet equation \eq{LOFNS}. 

The ratio
\begin{equation}
r_{NS,G}\equiv {G_{NS}(x,Q^2)\over G(x,Q^2)}\label{rNS-S},
\end{equation}
should vanish for solutions of the DGLAP equations, and any deviations are scaled to the size of the actual (singlet) gluon distribution $G_{\rm CTEQ5L}$. 

We consider specifically the non-singlet  combination of massless $u$ and $d$ quarks,
\ba 
F_{NS,ud}(x,Q^2)&\equiv &x\left[u(x,Q^2)+\bar u(x,Q^2)\right]\nonumber\\
&&-d(x,Q^2)-\bar d(x,Q^2)\left.\right], \label{FNSud}
\ea 
and calculate $r_{NS,G}$ for that combination, shown in the right-hand column of Fig. \ref{fig:doveru}. 
Any failure of the ratio to vanish indicates an inconsistency of the $u$ and $d$ distributions as solutions of the DGLAP evolution equations, and {\em absolute} magnitudes of the ratio near $1$ would indicate that the integrated discrepancies are large.

As seen from the right-hand column in Fig. \ref{fig:doveru}, the ratio $r_{NS,G}$ is zero for small $x$, but has large absolute values for $x\ga 0.1$ for $Q^2=5$, 20, and 100 GeV$^2$,  Figs. \ref{fig:doveru}\,(d), (e), and (f) respectively. In particular, around $x\sim 0.3$, the ratio is about unity, indicating a serious discrepancy near the maxima of the valence-quark distributions.

The pattern in $x$ of the discrepancies between the gluon distributions found analytically from either $F_2^{\gamma p}$ or $F_S$ and the distributions published  by the CTEQ5 group closely follows the pattern seen in the departure of ${\cal F}{\cal F}_{NS}$ from zero. For example, ${\cal F}{\cal F}_{NS}$ and the differences between the analytic and fitted gluon distributions all become negative or positive in the same regions. This   would seem to indicate that the quark distributions at large $x$ (and, in particular, the dominant $u$ and $d$ valence distributions) are the origin of the difficulty, the combination of their shapes and $Q^2$ dependence not being compatible with the DGLAP evolution equations, {\em either in } LO {\em or in}  NLO  in the $\overline {\rm MS}$ renormalization scheme. 

 
\section{ Conclusions}\label{sec:conclusions}

\begin{enumerate} 
\item
We have first developed a powerful  method for the analytic solution of the LO DGLAP evolution equations based on Laplace transforms.  This method allows us to determine $G(x,Q^2)$ directly from $F_2^{\gamma p}(x,Q^2)$ or other measurable structure functions such as $F_{2 (3)}^{\gamma Z}$, $F_{2 (3)}^Z$, and $F_{2 (3)}^{W^\mp}$, providing  close and simple connections to experiment for massless or effectively massless quarks. 

We can also determine $G(x,Q^2)$ analytically from the singlet quark distribution $F_S(x,Q^2)$ if the quark distributions are known, or $F_S(x,Q^2)$ from $G(x,Q^2)$ when the latter is known. The set of relations provide consistency checks on LO quark and gluon distributions obtained in other ways. These are the principal theoretical results of the paper.

\item As an illustration of our methods, we have used the analytic solutions to the evolution equations to obtain tests of the consistency of published quark and gluon distributions. In particular, we compare the gluon distributions determined analytically from  structure functions $F_2^{\gamma p}(x,Q^2)$ and $F_S(x,Q^2)$ calculated from the quark distributions determined in various analyses, to the gluon distributions given by the same analyses. We have found no problems for analytic and fitted gluon distributions for the MRST2001 LO or CTEQ6L solutions, while
the analytic and fitted gluon distributions for  CTEQ5, though consistent at small $x$, show small, but significant, deviations in the large $x$ region, indicating that  the quark and gluons distributions are not completely consistent \cite{codecomparisons}.
\item 
A further analysis using the non-singlet structure function $F_{NS}(x,Q^2)$ and both the LO CTEQ5L and the NLO CTEQ5M quark distributions showed definitively that there are problems with those distributions at high $x$. The quark distributions are not consistent with either LO (CTEQ5L) or NLO perturbative QCD in the $\overline{\rm MS}$ scheme (CTEQ5M) in the sense that they do not satisfy the appropriate non-singlet relations for $x\greaterabout  0.05$ even though they satisfy them very well at small $x$. The discrepancies, of unknown origin, are large for $x\sim 0.3$, and could  have serious consequences for predictions of processes sensitive to that $x$ region. We conclude that the CTEQ5 distributions \cite{CTEQ5} should not be used to make such predictions. Again, we found no problems with the MRST2001 distributions \cite{MRST2001} or the more recent CTEQ6 distributions \cite{CTEQ6,cteq6.5}.

\item We suggest that  groups that calculate parton distributions numerically from the coupled DGLAP equations  construct the appropriate {\em analytic solutions} to test the consistency of their results. 
\end{enumerate}

We note, finally, that we are working on an extension of our analytic LO solution for $G(x,Q^2)$ to include massive $c$ and $b$ quarks, using the methods of \cite{cteq6.5}, and on the NLO gluon solution.  


\begin{acknowledgments}
The authors would  like to thank the Aspen Center for Physics for its hospitality during the time much of this work was done. 

 D.W.M. receives support from DOE Grant No. DE-FG02-04ER41308.
 \end{acknowledgments}
 


\begin{thebibliography}{99}
%

\bibitem{bdm1} 
M.~M.~Block, L.~Durand, and D.~W.~McKay, Phys. Rev. D {\bf 77 }, 094003 (2008) [arXiv:0710.3212 [hep-ph]].

\bibitem{dglap} 

V.~N.~Gribov and L.~N.~Lipatov, Sov.~J.~Nucl.~Phys.~{\bf 15}, 438 (1972);  
G.~Altarelli and G.~Parisi, Nucl.~Phys.~{\bf B126}, 298 (1977);  
Yu.~L.~Dokshitzer, Sov.~Phys.~JETP~{\bf 46}, 641 (1977). 



\bibitem{CTEQ5} 
CTEQ Collaboration, H. L. Lai et al., Eur. Phys. J. {\bf C12}, 375 (2000) [hep-ph/9903282]. The LO CTEQ5L parton distributions were calculated using the Mathematica package CTEQ5L of the CTEQ group, http://www.phys.psu.edu/~cteq/Mathematica, and independently using the CTEQ5  distributions from the Durham parton distribution generator at http://durpdg.dur.ac.uk/hepdata/pdf3.html.  The parameters for the LO $\alpha_s(Q^2)$ are given in the introductory notes in the CTEQ5L Fortran program on the CTEQ web site, http://www.phys.psu.edu/~cteq/. 

\bibitem{MRST2001}
A.D. Martin, R.G. Roberts, W.J. Stirling and R.S. Thorne, MRST2001, Eur. Phys. J. {\bf C23},   73 (2002) [hep-ph/0110215].  The MRST2001 distributions were obtained using the Durham parton distribution generator at http://durpdg.dur.ac.uk/hepdata/pdf3.html.

\bibitem{CTEQ6}
J. Pumplin, D.R. Stump, J. Huston, H.L. Lai, P. Nadolsky, W.K. Tung, JHEP {\bf 02}, 07:012 (2002) [hep-ph/0201195]. 

\bibitem{
Floratos} E. G. Floratos, C. Kounnas and R. Lacaze, Nucl. Phys. {\bf B192}, 417 (1981).

\bibitem
{cteq6.5} CTEQ Collaboration, W. K. Tung, H. L. Lai,  A. Belyaev,  J. Pumplin,  D. Stump, and C.-P. Yuan, J. High Energy Phys. 0702:053, (2007) [hep-ph/0611254].

\bibitem{alphacteq} 
arXiv:hep-ph/0512167 v4, J. Pumplin et al. (2006); http://www.phys.psu.edu/~cteq/CTEQ5Table.

\bibitem{MRSTweb}
http://durpdg.dur.ac.uk/hepdata/mrs.html.

\bibitem{codecomparisons}
There were some hints that early  CTEQ pdf's might be inconsistent with those derived using other evolution codes; see J. Blumlein et al., hep-ph/9609400 (1996).

\end{thebibliography}
\end{document}